\documentclass[aps,prl,twocolumn,superscriptaddress]{revtex4}
\usepackage{graphicx}

\begin{document}

\title{Electronic structure in one-Fe Brillouin zone of iron-pnictide superconductor CsFe$_2$As$_2$}

\author{S. Kong}
\affiliation{National Synchrotron Radiation Laboratory, University of Science and Technology of China, Hefei, Anhui 230029, P. R. China}

\author{D. Y. Liu}
\email{dyliu@theory.issp.ac.cn}
\affiliation{ Key Laboratory of Materials Physics, Institute of Solid State Physics, Chinese Academy of Sciences, Hefei 230031, China}

\author{S. T. Cui}
\affiliation{National Synchrotron Radiation Laboratory, University of Science and Technology of China, Hefei, Anhui 230029, P. R. China}

\author{S. L. Ju}
\affiliation{National Synchrotron Radiation Laboratory, University of Science and Technology of China, Hefei, Anhui 230029, P. R. China}

\author{A. F. Wang}
\affiliation{Hefei National Laboratory for Physical Sciences at Microscale and Department of Physics, University of Science and Technology of China, Hefei, Anhui 230026, P. R. China}

\author{X. G. Luo}
\affiliation{Hefei National Laboratory for Physical Sciences at Microscale and Department of Physics, University of Science and Technology of China, Hefei, Anhui 230026, P. R. China}
\affiliation{Collaborative Innovation Center of Advanced Microstructures, Nanjing 210093, P. R. China}

\author{L. J. Zou}
\affiliation{ Key Laboratory of Materials Physics, Institute of Solid State Physics, Chinese Academy of Sciences, Hefei 230031, China}

\author{X. H. Chen}
\affiliation{Hefei National Laboratory for Physical Sciences at Microscale and Department of Physics, University of Science and Technology of China, Hefei, Anhui 230026, P. R. China}
\affiliation{Collaborative Innovation Center of Advanced Microstructures, Nanjing 210093, P. R. China}

\author{G. B. Zhang}
\affiliation{National Synchrotron Radiation Laboratory, University of Science and Technology of China, Hefei, Anhui 230029, P. R. China}
\affiliation{Collaborative Innovation Center of Advanced Microstructures, Nanjing 210093, P. R. China}

\author{Z. Sun}
\email{zsun@ustc.edu.cn}
\affiliation{National Synchrotron Radiation Laboratory, University of Science and Technology of China, Hefei, Anhui 230029, P. R. China}
\affiliation{Collaborative Innovation Center of Advanced Microstructures, Nanjing 210093, P. R. China}





\maketitle

\textbf{The multiband nature of iron-pnictide superconductors is one of the keys to the understanding of their intriguing behavior \cite{PJH}. The electronic and magnetic properties heavily rely on the multiband interactions between different electron and hole pockets near the Fermi level. At the fundamental level, though many theoretical models were constructed on the basis of the so-called 1-Fe Brillouin zone (BZ) with an emphasis of the basic square lattice of iron atoms \cite{PJH, Raghu, PALee, KurokiPRL, Kemper, Graser, Daghofer,Kuroki, Anderson,Brouet}, most electronic structure measurements were interpreted in the 2-Fe BZ \cite{DHLu, Kondo, TSato, Zabolotnyy, PRichard, TYoshida, MYi, YZhangPRB, YZhang, DLiu}. Whether the 1-Fe BZ is valid in a real system is still an open question. Using angle-resolved photoemission spectroscopy (ARPES), here we show in an extremely hole-doped iron-pnictide superconductor CsFe$_2$As$_2$ that the distribution of electronic spectral weight follows the 1-Fe BZ, and that the emerging band structure bears some features qualitatively different from theoretical band structures of the 1-Fe BZ. Our analysis suggests that the interlayer separation is an important tuning factor for the physics of FeAs layers, the increase of which can reduce the coupling between Fe and As and lead to the emergence of the electronic structure in accord with the 1-Fe symmetry of the Fe square lattice. Our finding puts strong constraints on the theoretical models constructed on the basis of the 1-Fe BZ. }

The generic crystallographic structure of iron-pnictides is the Fe-As plane with As atoms above and below the square lattice of Fe atoms (see Fig. 1a). The alternating pattern of As makes the nearest neighboring Fe atoms inequivalent and leads to a large crystallographic unit cell with two Fe atoms. For 122 family of iron-pnictides, the three-dimensional Brillouin zone is shown in Fig. 1b. In real space, the 2-Fe unit cell is twice as large as the unit cell of Fe square lattice, the corresponding so-called 2-Fe BZ in momentum space is half the size of the 1-Fe BZ based on the Fe square lattice. In iron-pnictides, the 2-Fe BZ reflects the full symmetry of the crystallographic and electronic structures \cite{PJH, Singh, Boeri, Ma, DHLu, Kondo, TSato, Zabolotnyy, PRichard, TYoshida, MYi, YZhangPRB, YZhang, DLiu}.

On the other hand, many theoretical models of band structures have been developed in the 1-Fe BZ for its simplicity \cite{PJH, Raghu, PALee, KurokiPRL, Kemper, Graser, Daghofer,Kuroki, Anderson,Brouet}, and the importance of the fundamental Fe square lattice has been suggested by some experiments \cite{JTPark,Lumsden,HFLi}. Though the 2-Fe BZ should be adopted for describing the full symmetry of iron-pnictides, the 1-Fe BZ sometimes comes with convenience. More importantly, if the spectral weight distribution follows the 1-Fe BZ instead of the 2-Fe BZ, the electronic excitations in momentum space can be highly different. From the perspective of 1-Fe unit cell, the As anions break the translational symmetry of the primary Fe square lattice and introduce ``shadow" bands in the momentum space by folding the main bands with a reciprocal vector. When the coupling between the As anions and the Fe square lattice becomes stronger, the spectral weight of the ``shadow" bands will increase accordingly. The intensity of shadow bands in most iron-pnictide materials is comparable with that of the main bands, which results in the electronic spectral weight distribution in accord with the 2-Fe BZ in the electronic structure measurements of ARPES. Recently, Li \emph{et al.} attempted to recover the spectral weight distribution in the 1-Fe BZ through unfolding the band structures in the 2-Fe BZ from the perspective of first-principles calculations  \cite{CLin}.

Indeed, if the shadow band intensity is weakened, the main bands of the 1-Fe BZ can be revealed and uncover fundamental interactions in the Fe square lattice. However, the strong intensity of shadow bands in most iron-pnictide materials makes the main bands of the 1-Fe BZ indistinguishable, and a direct manifestation of the electronic spectral weight distribution in the 1-Fe BZ has not been realized experimentally. Using ARPES, here we report the electronic structures of a superconductor CsFe$_2$As$_2$ (T$_c$ = 1.8 K). Contrary to other iron pnictides, for the first time we are able to demonstrate the primary spectral weight distribution (or the main bands) in the 1-Fe BZ, which suggests that the shadow bands induced by As anions in  CsFe$_2$As$_2$ is rather weak. Our theoretical calculations suggest that this unusual behavior is induced by the increased separation between FeAs layers in CsFe$_2$As$_2$, which reduces the Fe-As coupling along the \textit{c}-axis and weakens the folding potential originating from the alternating spatial configuration of As atoms. Our data reveals a realistic band structure of the 1-Fe BZ, which puts strong constraints on the theoretical models constructed on the basis of the 1-Fe BZ.

\begin{figure}
\includegraphics[scale=0.42]{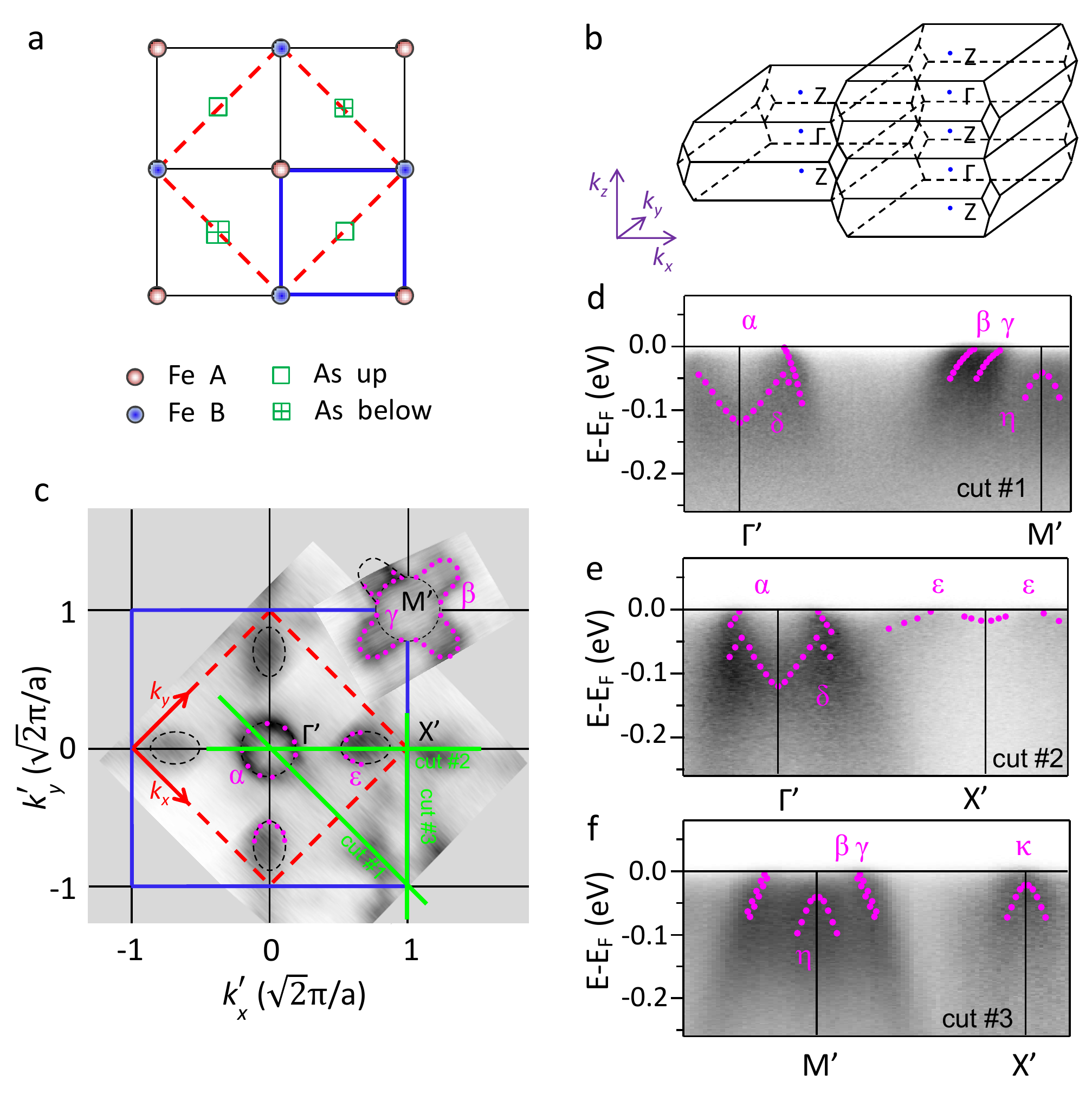}
\caption{\textbf{Fermi surface topology and near-E$_F$ bands of CsFe$_2$As$_2$}. \textbf{a}, A sketch of Fe-As layer with As anions located above and below the Fe plane. The blue thick lines show the 1-Fe unit cell of the primary Fe square lattice. The red dashed lines define the crystallographic two-dimensional unit cell with inequivalent Fe at the corners and the center. Their corresponding Brillouin zones in the momentum space are referred to as the 1-Fe BZ and 2-Fe BZ, respectively. \textbf{b}, The Brillouin zone of 122 iron-based superconductors. \textbf{c}, Photoemission intensity mapping at the Fermi energy for CsFe$_2$As$_2$ taken with 85 eV and circular polarized photons. The blue and red squares illustrate the 1-Fe BZ and 2-Fe BZ, respectively. The dots indicate the $k_F$ positions determined by the peak positions of the momentum distribution curves, and the thin dashed lines show the Fermi surface pockets deduced from the  $k_F$ positions. In the 1-Fe BZ, there are one hole Fermi surface sheet around the zone center $\Gamma$$'$, one hole sheet near the middle point of the zone boundary X$'$ or Y$'$, one flower-like and one circular hole pockets around the zone corner M$'$. \textbf{d}-\textbf{f}, Photoemission intensity plots along cut $\#$1, cut $\#$2 and cut $\#$3 in panel \textbf{c}, respectively. The dotted lines are guides for the eyes to trace the dispersions of near-E$_F$ electronic bands.}
\label{Fig1}
\end{figure}

Fig. 1c shows a typical mapping of spectral weight distribution at the Fermi energy in CsFe$_2$As$_2$, and the intensity plots along three high-symmetry cuts are displayed in Figs. 1d-f. Photon energies have been varied to reach different $k_z$ values in momentum space, and similar measurements have also been performed on RbFe$_2$As$_2$, which all demonstrate a consistent spectral weight distribution (see the supplementary materials). We thus believe that Fig. 1c shows a generic Fermi surface topology of CsFe$_2$As$_2$ (also for RbFe$_2$As$_2$) in the $k_x$-$k_y$ plane with rather weak dispersion in $k_z$ direction. 

In Fig. 1c, we sketch the zone areas for both the 1-Fe (blue) and the 2-Fe (red) BZs, and we use prime symbol to mark the high symmetry points in the 1-Fe BZ. Here we would like to point out the spectral weight distribution matches the 1-Fe BZ, in contrast to the 2-Fe BZ that has been widely observed in ARPES measurements on most iron-pnictide materials. In the 1-Fe BZ, there are one hole Fermi surface sheet around the zone center $\Gamma'$, one hole sheet near the middle point of the zone boundary X$'$ or Y$'$, one flower-like and one circular hole pockets around the zone corner M$'$. In the 2-Fe BZ, both $\Gamma'$ and M$'$ are located along the $\Gamma$-Z-$\Gamma$ high symmetry lines in the $k_z$ direction. Within the 2-Fe BZ, the flower-like hole pocket around M$'$ is also expected to appear around the zone center for some  $k_z$ values, which however is against our Fermi surface mappings with various photon energies (see the supplementary materials for more details). Our data shows that the band structures around $\Gamma'$ and M$'$ are inequivalent, and the 1-Fe BZ is more suited to the distribution of the spectral weight.

We stress that the band structures follow the 2-Fe BZ. However, from the perspective of the 1-Fe BZ, what we have shown in Fig. 1c should be regarded as the spectral weight distribution of main bands in the 1-Fe BZ, which emerges when the shadow bands induced by As anions bear weak intensity. The visibility of main bands in the 1-Fe BZ provides an opportunity to examine the band structures that are more relevant to the Fe square lattice.

The band dispersions of CsFe$_2$As$_2$ in the 1-Fe BZ can be manifested by the photoemission intensity plots along cuts $\#$1$-$$\#$3, as shown in Figs. 1d-f, respectively. The hole-like band, $\alpha$, forms the circular Fermi surface sheet around the zone center $\Gamma$$'$. Near the zone corner M$'$, two hole-like bands, $\beta$ and $\gamma$, can be resolved. They construct the large flower-shaped Fermi surface and the small circular one (see Fig. 1c). Near the X$'$ point, there is a hole-like pocket $\varepsilon$. In addition, there are $\delta$, $\eta$ and $\kappa$ bands around $\Gamma$$'$, M$'$ and X$'$, respectively, staying below the Fermi level. Along cut $\#$3, $\beta$ and $\gamma$ are nearly degenerate and cannot be clearly decoupled.

\begin{figure}
\includegraphics[scale=0.58]{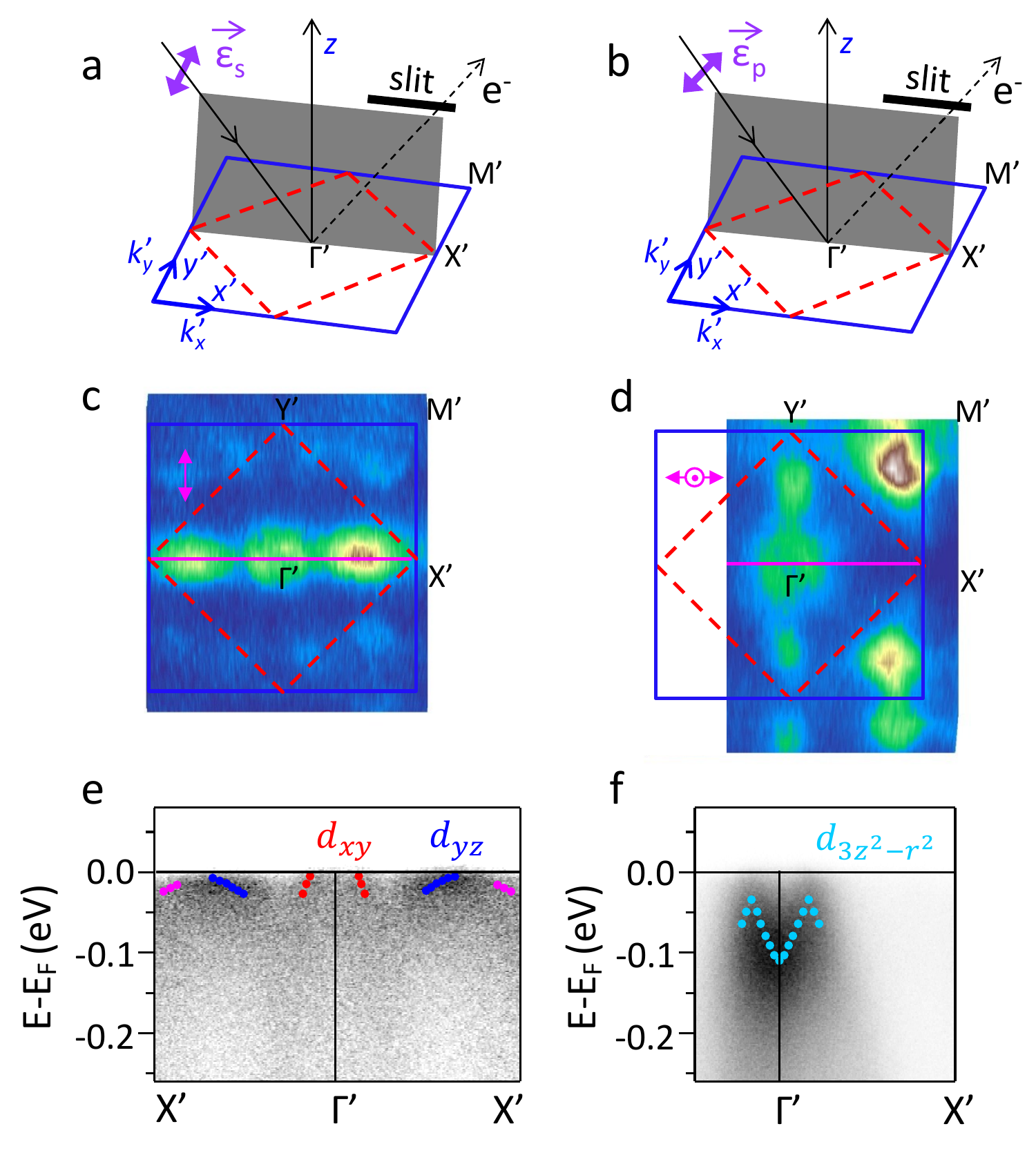}
\caption{\textbf{Orbital characters of near-E$_F$ electronic bands}. \textbf{a},\textbf{b}, The experimental setups with different photon polarizations. The blue solid lines and red dashed lines indicate the orientations of the 1-Fe and 2-Fe BZs, respectively. The \textit{x}$'$, $k_x$$'$, \textit{y}$'$ and $k_y$$'$ axes are along the neighboring Fe-Fe directions. The grey planes are mirror planes for the analyses of orbital symmetries. \textbf{c},\textbf{d}, The experimental Fermi surface mapping using the setups in panels \textbf{a} and \textbf{b}, respectively. All data were taken using 85 eV photons at T = 25 K. The blue solid lines and red dashed lines illustrate the 1-Fe BZ and 2-Fe BZ, respectively. The pink arrows indicate the light polarizations. \textbf{e},\textbf{f}, Photoemission intensity plots taken along the $\Gamma$$'$-X$'$ cuts in panels \textbf{c} and \textbf{d}, respectively. }
\label{Fig2}
\end{figure}

In iron pnictides the near-E$_F$ electronic bands consist primarily of Fe 3\textit{d} states of $d_{xy}$,  $d_{xz}$, $d_{yz}$, $d_{x^2-y^2}$ and $d_{3z^2-r^2}$  orbitals. These orbital characters can be resolved by varing the polarization of photons in ARPES measurements \cite{MYi, YZhangPRB, YZhang}. Figs. 2a and 2b show the experimental setups with primary \textit{s} and \textit{p} light polarizations, with the orientations of the 1-Fe and 2-Fe BZs illustrated. Figs. 2c and 2d show the spectral weight distribution at the Fermi level taken with \textit{s} and \textit{p} light polarizations, respectively. Figs. 2e and 2f display the corresponding photoemission intensity taken along the mirror planes (along the $\Gamma$$'$-X$'$ cuts in Figs. 2c and 2d), respectively. By analyzing the spectral weight distribution for different experimental setups, we can derive the orbital characters of individual bands (see the supplementary materials for more details).

\begin{figure}
\includegraphics[scale=0.5]{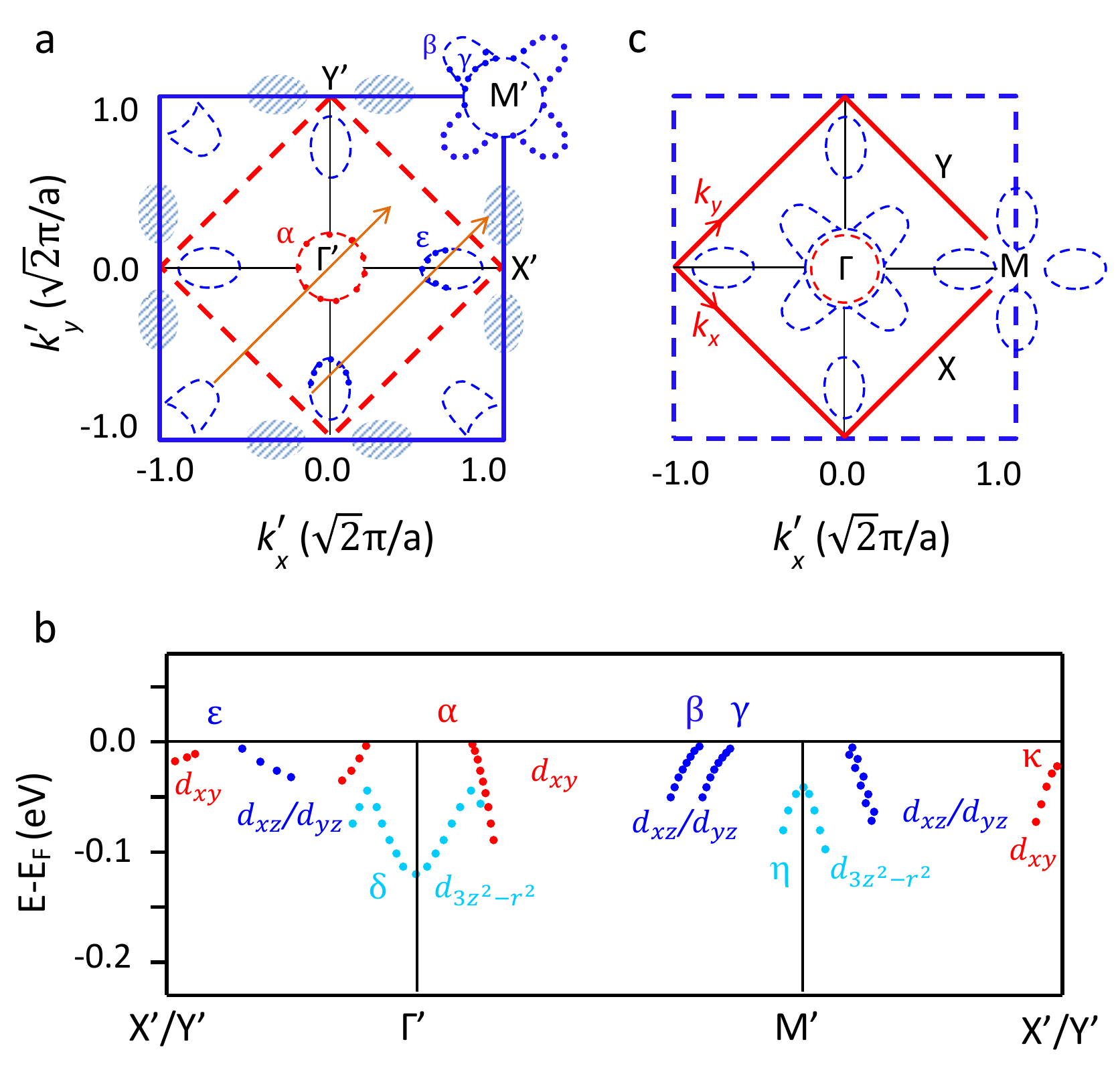}
\caption{\textbf{A summary of electronic structures of CsFe$_2$As$_2$}. \textbf{a}, The Fermi surface of CsFe$_2$As$_2$ determined by experimental data. The blue and red squares represent the 1-Fe BZ and 2-Fe BZ, respectively. The hatched areas indicate the locations of the $\varepsilon$ pockets in theoretical models based on the 1-Fe BZ. \textbf{b}, The band dispersions of CsFe$_2$As$_2$ in the 1-Fe BZ, determined by Figs. 1d-f. \textbf{c}, The constructed Fermi surface in the 2-Fe BZ by folding all bands with a vector of ($\pi$,$\pi$) of the 1-Fe BZ. The folding vector is indicated in panel \textbf{a}.   }
\label{Fig3}
\end{figure}

The Fermi surface sheets, band dispersions and orbital characters derived from experimental data are summarized in Figs. 3a and 3b, respectively.
On the basis of the 1-Fe BZ, the folding potential induced by As atoms leads to a shifting of all bands with a vector of ($\pi$,$\pi$,$\pi$) to reconstruct the band structure in the 2-Fe BZ. Owing to the weak $k_z$ dispersion, we choose a vector of  ($\pi$,$\pi$) to perform this reconstruction in the $k_x$-$k_y$ plane. Using the spectral weight distribution in the 1-Fe BZ, we plot the Fermi surface topology in the 2-Fe BZ in Fig. 3c, which shows three hole-like Fermi surfaces ($\alpha$, $\beta$, $\gamma$) around the zone center and four small hole-like pockets ($\varepsilon$) near the zone corners. In Fig. 3c, we estimate that the hole-like Fermi surface sheets of $\alpha$, $\beta$, $\gamma$ and $\varepsilon$ occupy approximately 6\%, 22\%, 7\%, 12\% of the 2-Fe BZ, respectively, which yields a total of $\sim$ 47 \% of this area. This value is in excellent agreement with the nominal hole-doping of 50 \% in CsFe$_2$As$_2$. 

It is a crucial concern to determine a realistic electronic structure in momentum space. Comparing Figs. 3a and 3c, one can realize that the reconstructed band structure, similar to the band structures in the 2-Fe BZ of many iron-pnictide materials, mixes individual features of both the main bands and shadow bands. Such a mixture makes it difficult to resolve the characteristics in the main bands. Moreover, since the electronic excitations highly rely on the spectral weight distribution in momentum space, the choice of the 1-Fe or 2-Fe BZs may result in a variation of theoretical conclusions. In the present study, the special spectral weight distribution outlines the dominant dispersive bands of CsFe$_2$As$_2$ in the 1-Fe BZ. Though a theoretical attempt to construct the spectral weight distribution in the 1-Fe BZ has been made for some iron pnictides on the basis of band structures in the 2-Fe BZ \cite{CLin}, our data do not match the generic band structures as predicted in Ref. 27.

We have performed band calculations for CsFe$_2$As$_2$ in the 2-Fe BZ using the full potential linearized augmented plane-wave scheme (see the supplementary materials). Though such calculations generally achieve an excellent agreement with ARPES measurements for most iron-pnictide materials, we find that the calculated Fermi surface topology for CsFe$_2$As$_2$ is quantitatively different from the reconstructed one shown in Fig. 3c. This distinction suggests significant correlation effects in CsFe$_2$As$_2$, which have not been caught in the simple LDA calculations. 

A direct comparison of Fermi surface topologies in ARPES data and the theoretical band structures in the 1-Fe BZ reveals two distinct features. First,  our data shows one Fermi surface pocket of $d_{xy}$ orbital symmetry at the zone center, while theoretical models suggest that either one pocket with $d_{xy}$ symmetry or two pockets with $d_{xz}$/$d_{yz}$ symmetry can exist around the zone center of the 1-Fe BZ  \cite{PJH, Raghu, PALee, KurokiPRL, Kemper, Graser, Daghofer,Kuroki, Anderson,Brouet}. Owing to the use of glide-mirror symmetry, the latter possibility is ruled out. Second, in the scenario of one $d_{xy}$ pocket around the zone center, theoretical calculations consistently suggest that the $\varepsilon$ pockets should appear in the hatched regions of Fig. 3a, instead of their current locations in our data.  As a matter of fact, the strong spectral weight of $\varepsilon$ pockets in our data emerges in the ``shadow" bands rather than in the main bands (the hatched regions). This anomaly could be relevant to some unusual ``shadow" band effects, suggesting the Fe-As coupling still plays an important role to affect the spectral weight distribution. The disparities between experimental data and theory put strong constraints on a realistic band structure of theoretical models in the 1-Fe BZ. Moreover, in heavily hole-doped systems, such as CsFe$_2$As$_2$ and RbFe$_2$As$_2$, the correlation effects are very strong \cite{Werner}. It is highly interesting to investigate whether such correlation effects could play a critical role for the remarkable differences between our data and theoretical band structures in the 1-Fe BZ.

In the physics of iron-pnictides, As anions play crucial roles. For instance, the position of As above and below the Fe square lattice is closely related to the critical transition temperatures as well as the superconducting order parameters \cite{Kuroki,Mizuguhci,HOkabe,KHashimoto}. Moreover, it has also been proposed recently that there is odd-parity superconductivity in these materials due to the spacial configuration of As \cite{JHu}.
The presence of As provides a translational symmetry broken potential that can induce a band folding from the 1-Fe BZ into the 2-Fe BZ, though the lack of evident spectral weight of the folded bands in CsFe$_2$As$_2$ suggests that the folding potential of As is substantially reduced in this material. 
We argue that the interlayer coupling between FeAs layers plays a key role to the understanding of the unusual Fermi surface topology of CsFe$_2$As$_2$. The large separation of FeAs layers along the \textit{c}-axis weakens the interlayer coupling in CsFe$_2$As$_2$. Such a large separation makes the system more two-dimensional and enhances the electronic correlations. Accordingly, the Fe 3\textit{d} electrons become localized in the \textit{c}-axis, and the hybridization between Fe 3\textit{d} and As 4\textit{p} orbitals greatly reduces in the \textit{c}-axis direction (see the supplementary materials). This reduction can considerably decrease the folding potential of As and lead to the emergence of spectral weight distribution in the 1-Fe BZ in CsFe$_2$As$_2$.  From this perspective, the interlayer coupling between FeAs layers is a particularly interesting factor for tuning the physics of FeAs layer itself. In addition to CsFe$_2$As$_2$ and RbFe$_2$As$_2$, we expect more iron-pnictide compounds with a large \textit{c}-axis separation can demonstrate similar 1-Fe BZ features.  \\

\textbf{Methods}

 High-quality single crystals of CsFe$_2$As$_2$ were grown using self-flux method. Superconductivity was observed in these crystals with T$_c$=1.8K. ARPES experiments were performed at the SIS beamline of the Swiss Light Source (SLS), the beamline 4.0.3 of the Advanced Light Source, the BL7U of the UVSOR, and the beamline 28A of the Photon Factory (PF), using Scienta R4000 and SES-2002  electron spectrometers, respectively. The angular resolution was 0.3 degrees and the combined instrumental energy resolution was better than 20 meV.  The polarization of the photons was varied to investigate the band dispersions and orbital characters. All samples were cleaved and measured at 25 K under a vacuum better than $1\times 10^{-10}$ mbar. \\
 
 \textbf{Acknowledgements}
 The authors are grateful for the discussions with Dan Dessau, Donglai Feng, Jiangping Hu and Tao Wu, and the experimental support by M. Shi, N. C. Plumb, N. Xu at the SLS, J. D. Denlinger at the ALS, and K. Ono, Y. Ishida at the  Photon Factory, and M. Matsunami and S. Kimura at the UVSOR, and Q. Fan from the Fudan University. This work was supported by National Natural Science Foundation of China, the National Basic Research Program of China (973 Program, Grant No. 2012CB922004, 2014CB921102, 2012CB922002), the Chinese Academy of Sciences, the “Strategic Priority Research Program (B)” (Grant No. XDB04040100). Z. S acknowledges the support by the Fundamental Research Funds for the Central Universities.



{\Large \textit{Supplementary Materials}}\\

\textbf{I. Sample preparation}

High quality CsFe$_{2}$As$_{2}$ single crystals were grown by the self flux technique. The Cs chunks, Fe and As powder were weighted with the ratio Cs:Fe:As=6:1:6. The mixture of Fe and As powders were loaded into an alumina crucible, and freshly cut Cs pieces were placed on top of the mixture. Then the alumina crucible with a lid was sealed in a stainless steel container. The whole preparation process was carried out in the glove box in which high pure argon atmosphere. The sealed stainless steel container was then sealed inside an evacuated quartz tube. The quartz tube was placed in a box furnace and slowly heated up to 200 $^\circ$C, and kept at 200 $^\circ$C for 400 minutes to ensure the reaction of Cs with the mixture. Then the sample was heated up to 950 $^\circ$C in 10 hours. The temperature was kept for 10 hours and then slowly cooled to 550 $^\circ$C at a rate of 3 $^\circ$C/h. After cooling down to room temperature , shiny single crystals can be found in the alumina crucible. The  RbFe$_{2}$As$_{2}$ single crystals were grown using the same procedures. Fig. 4 shows the in-plane resistivity as the function of temperature for CsFe$_{2}$As$_{2}$ and RbFe$_{2}$As$_{2}$ single crystals.\\

\textbf{II. Orbital symmetries}

 Generally speaking, using \textit{s}-polarized photons as shown in Fig. 2a, we can suppress the orbital states of even symmetry with respect to the mirror plane and retain the orbital states with odd symmetry. That is, the spectral weight of $d_{yz}$ orbital is stronger than that of $d_{xz}$, $d_{x^2-y^2}$ and $d_{3z^2-r^2}$ orbitals. We note here that, according to calculations, the intensity of $d_{xy}$ states is expected to be much weaker than that of $d_{yz}$ states, though the $d_{xy}$ shows odd symmetry with respect to the mirror plane shown in Fig. 2a \cite{MYi1, YZhangPRB1, YZhang1}. In Fig. 2e, the $\alpha$ pocket shows weak spectral weight, consistent with the $d_{xy}$ symmetry as suggested by theoretical models based on 1-Fe BZ \cite{Anderson1,Brouet1, MYi1, YZhang1}. A portion of the $\varepsilon$ pocket shows strong intensity, which is in agreement with the $d_{yz}$ character indicated by band calculations (see the Section IV). Moreover, the $\varepsilon$ band contains $d_{xy}$ states in the region that is close to the X$'$ point, and accordingly we observed weak spectral weight in this region. On the other hand, the \textit{p} polarized photons can suppress odd orbitals and retain even orbitals with respect to the mirror plane. Based on the spectral weight in Fig. 2f and band calculations (see the supplementary materials), one can realize that the $\delta$ band consists of $d_{3z^2-r^2}$ states. 
 
 \begin{figure}[htbp]\centering
 \includegraphics[angle=0, width=0.8 \columnwidth]{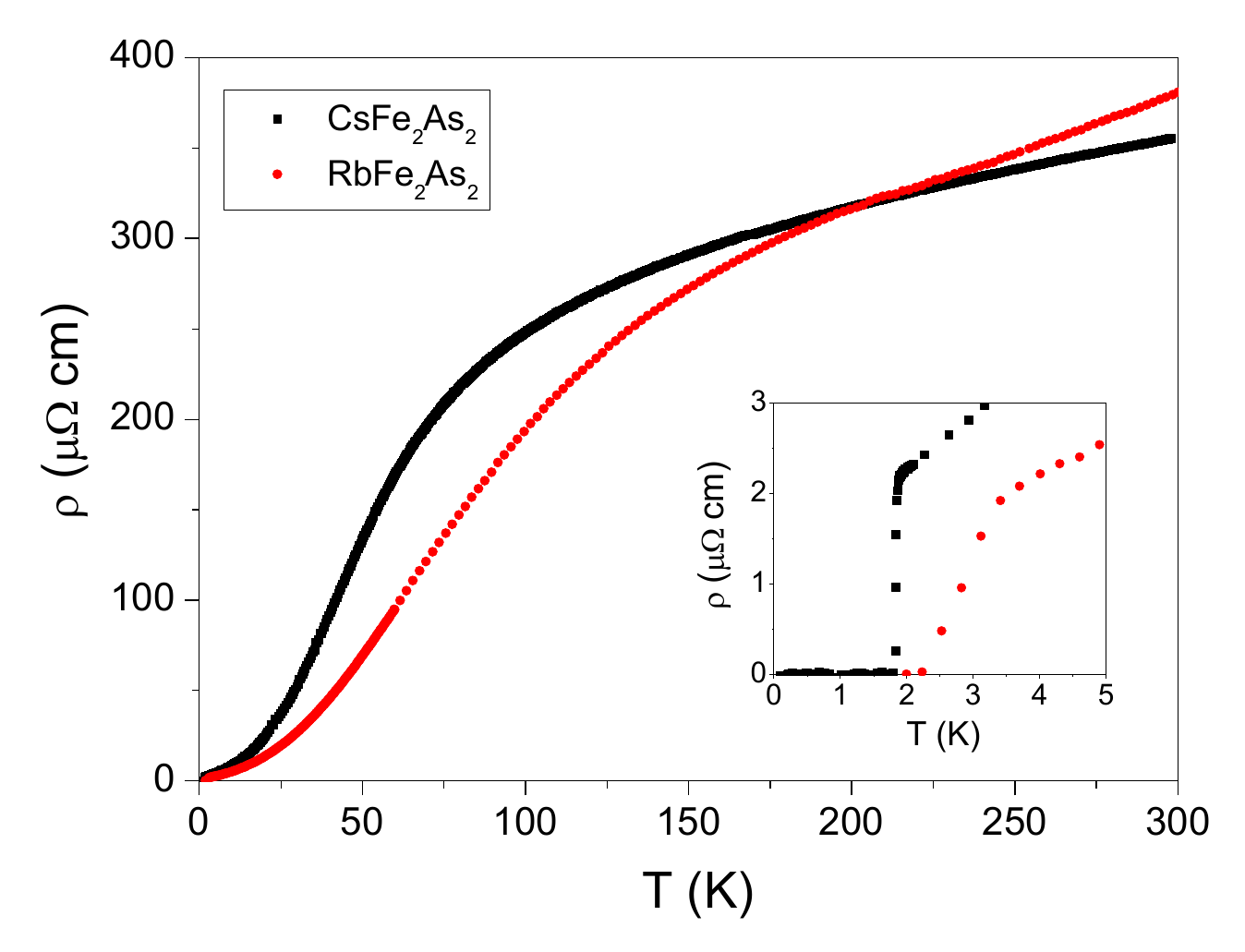}
 \caption{Resistivity plotted as a function of temperature for CsFe$_{2}$As$_{2}$ and RbFe$_{2}$As$_{2}$ single crystal. The inset is the zoom plot of resistivity around the superconducting transition.}
 \end{figure}
 
 The combination of light polarization and orbital symmetry can also affect the distribution of photoemission intensity away from the mirror planes. In Fig. 2c, the $\beta$ pocket disappears, while it shows up with strong spectral weight in Fig. 2d. In light of spectral weight distribution as shown in Refs. \cite{MYi1, YZhang1}, this discrepancy indicates that there are significant $d_{xz}$ states in the $\beta$ band. Moreover, the $\varepsilon$ pocket in the setup of \textit{s} polarization presents itself near X$'$ point and becomes invisible near Y$'$ point (Fig. 2c), while it emerges near Y$'$ and disappears near X$'$ in the \textit{p} polarization setup (Fig. 3d). This behavior indicates that the $\varepsilon$ pocket possesses strong $d_{yz}$ and $d_{xz}$ states near the X$'$ and Y$'$ points, respectively. Indeed, this argument still holds with the presence of twined structural domains in single crystals,  since the character of $d_{xz}$ is equivalent to that of $d_{yz}$ orbitals after 90$^\circ$ rotation due to the $C_4$ rotational symmetry. 
 
 The orbital character of the $\gamma$ band cannot be clearly resolved in our data, because it is very close to the $\beta$ band. Nevertheless, since theoretical models suggest that the two hole pockets of $d_{yz}$ and $d_{xz}$ orbitals always accompany each other around either $\Gamma$$'$ or M$'$ points in 1-Fe BZ \cite{PJH1, PALee1, Kemper1, Graser1, Kuroki1, Anderson1,Brouet1}, the $\gamma$ band should possess $d_{yz}$ and $d_{xz}$ orbital states. The orbital characters of the $\eta$ and $\kappa$ bands have not been clearly judged by the symmetry analysis. However, based on theoretical models in the 1-Fe BZ, we ascribe them to $d_{3z^2-r^2}$ and $d_{xy}$ orbital states, respectively \cite{Anderson1}.

\textbf{III. Electronic structure}

\begin{figure}[htbp]\centering
\includegraphics[angle=0, width=0.8 \columnwidth]{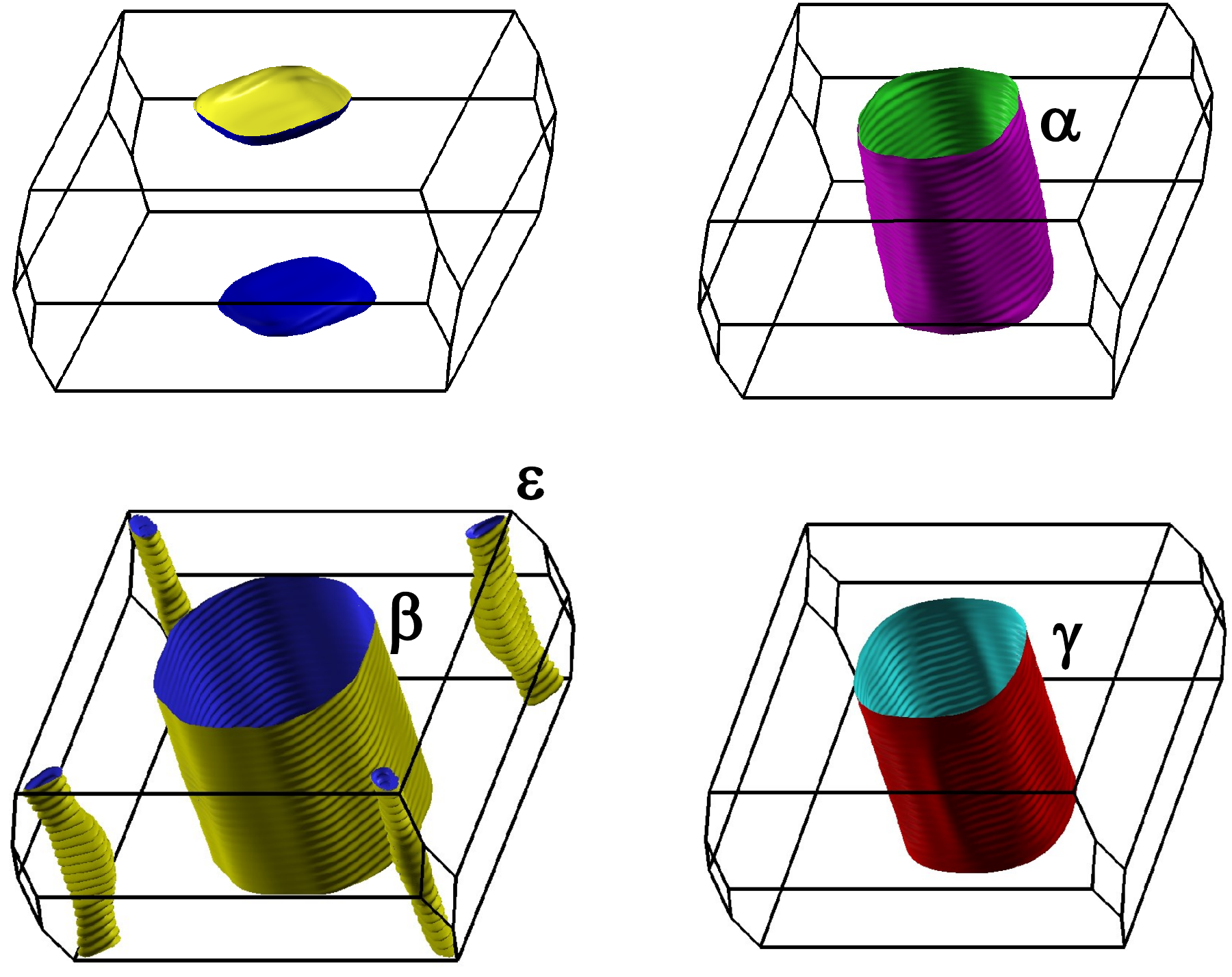}
\caption{Fermi surface sheets of CsFe$_{2}$As$_{2}$ obtained by LDA calculations.
} 
\label{figS2}
\end{figure}

The electronic structure calculations for CsFe$_{2}$As$_{2}$ in the 2-Fe BZ were performed using the full potential linearized augmented plane-wave (FP-LAPW) scheme in the WIEN2k programme package \cite{WIEN2K1}. The lattice parameters were adopted from the experimental data of CsFe$_{2}$As$_{2}$ \cite{ZAAC620-17771}. Notice that the significant difference of the lattice parameters between CsFe$_{2}$As$_{2}$ and its sister compound KFe$_{2}$As$_{2}$ is that the former has a more larger $c$ = 15.0997 \AA \cite{ZAAC620-17771} than that of the latter with $c$ = 13.861 \AA \cite{ZNB36-16681}, owing to the larger radius of Cs ions. Moreover, the As-Fe-As angle is 109.58$^{\circ}$ (indicating a nearly ideal tetrahedron) in CsFe$_{2}$As$_{2}$, while it is 107.03$^{\circ}$ in KFe$_{2}$As$_{2}$, implying different crystal field splittings. Based on the analysis of the obtained on-site energies, the splitting of the three 4$p$ orbitals of As is nearly zero in CsFe$_{2}$As, suggesting that the three 4$p$ orbitals of As are degenerate. On the contrary, a finite value about $-$0.12 eV between the $p_{x}$/$p_{y}$ and $p_{z}$ orbitals is found in KFe$_{2}$As$_{2}$. This difference could be relevant to different electronic structure between the sister compounds CsFe$_{2}$As$_{2}$ and KFe$_{2}$As$_{2}$.

\begin{figure}[htbp]\centering
\includegraphics[angle=0, width=1 \columnwidth]{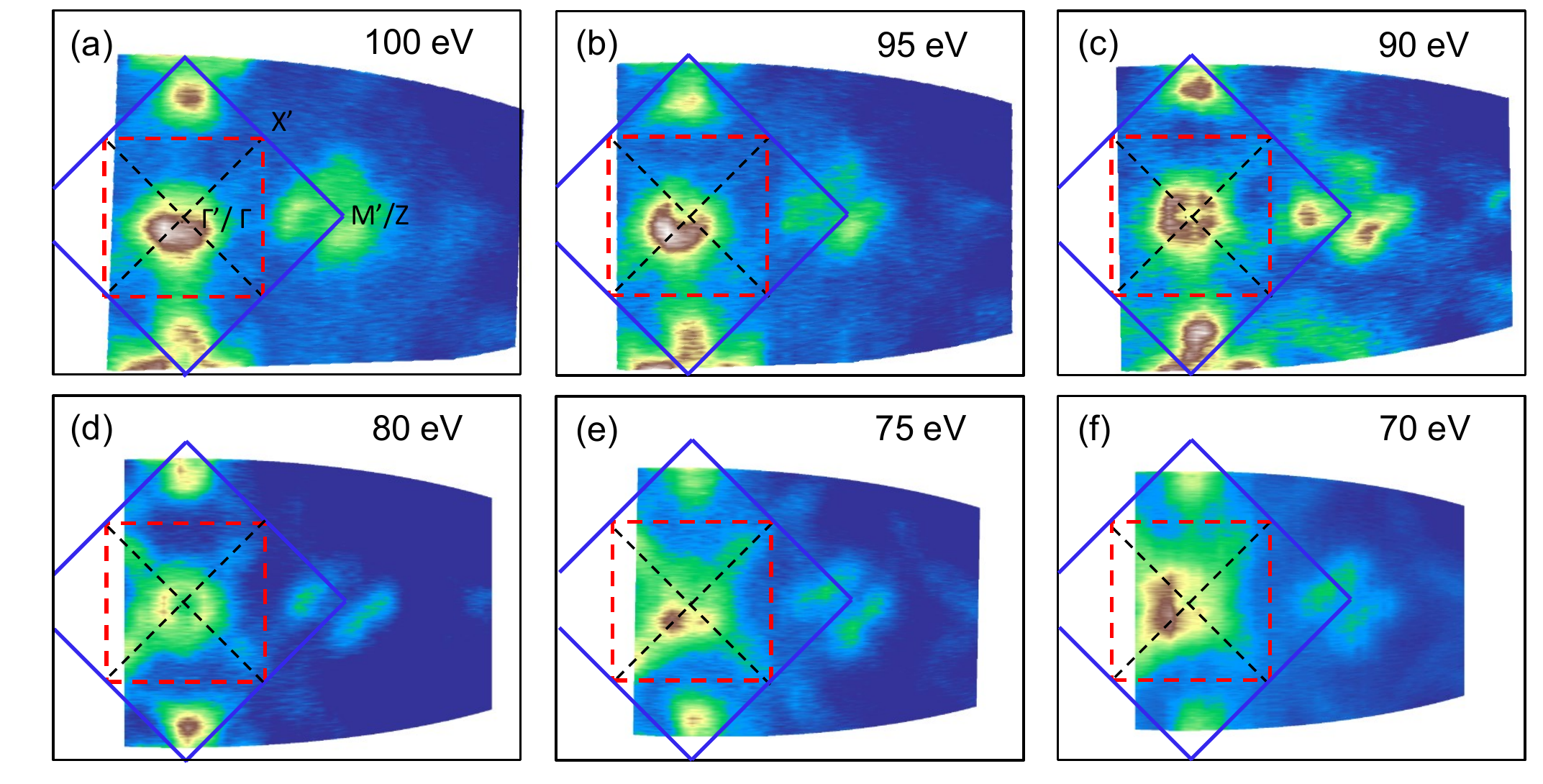}
\caption{Fermi surface mappings of CsFe$_{2}$As$_{2}$ obtained using various photon energies. The data were measured at T=25 K with circular-polarized photons. The blue solid lines and red dashed lines indicate the
orientations of the 1-Fe and 2-Fe BZs, respectively. High symmetry points in the 1-Fe BZ are marked with prime symbol.
} 
\label{fig3}
\end{figure}

The Fermi surface of CsFe$_{2}$As$_{2}$ was obtained within the local density approximation (LDA), as shown in Fig. 5. It consists three hole-like Fermi surface cylinders (denoted as $\alpha$, $\beta$ and $\gamma$ sheets) around the $\Gamma$ point, four hole-like pockets (denoted as $\varepsilon$) near the zone corner $M$ point, and one hole-like pocket around the $Z$ point.

\begin{figure}[htbp]\centering
\includegraphics[angle=0, width=1 \columnwidth]{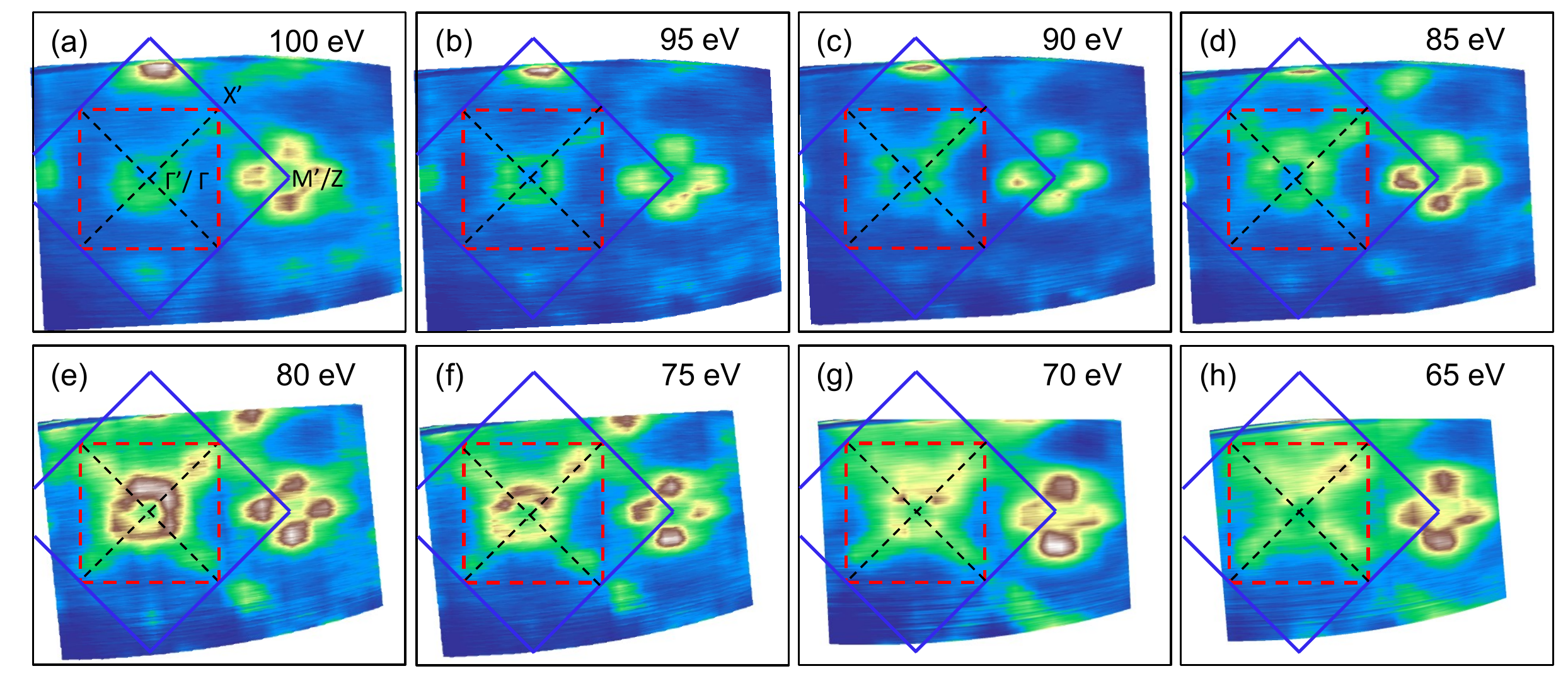}
\caption{Fermi surface mappings of RbFe$_{2}$As$_{2}$ obtained using various photon energies. The data were measured at T=25 K with circular-polarized photons. The blue solid lines and red dashed lines indicate the
orientations of the 1-Fe and 2-Fe BZs, respectively. High symmetry points in the 1-Fe BZ are marked with prime symbol.
} 
\label{fig4}
\end{figure}

LDA+DMFT calculations for KFe$_{2}$As$_{2}$ have suggested that the three Fermi surface sheets around $\Gamma$ point are renormalized \cite{nmat10-9321}. As a comparison, the three-dimensional character in CsFe$_{2}$As$_{2}$ is weaker than that in KFe$_{2}$As$_{2}$, and the electronic correlation and spin/orbital fluctuations in CsFe$_{2}$As$_{2}$ are expected to be stronger than those in KFe$_{2}$As$_{2}$. It is reasonable to believe that the three pockets around $\Gamma$ will also be renormalized when the electronic correlation and the spin/orbital fluctuations are taken into account. This indicates that the electronic correlation in CsFe$_{2}$As$_{2}$ will have obvious renormalizations in its band structures and unique FS topology. Indeed, our LDA+$U$ calculations showed that with increasing Coulomb interaction, a fractional of quasiparticle spectral weight shifts from the Fermi surface sheets around the $\Gamma$ point to those near the zone corner, the size of the $\varepsilon$ pockets becomes larger and the three pockets around $\Gamma$ point are renormalized to smaller ones. We thus believe that electronic correlations contribute to the difference between the Fermi surface topologies in our LDA calculations and ARPES measurements.

\begin{figure}[htbp]\centering
\includegraphics[angle=0, width=1 \columnwidth]{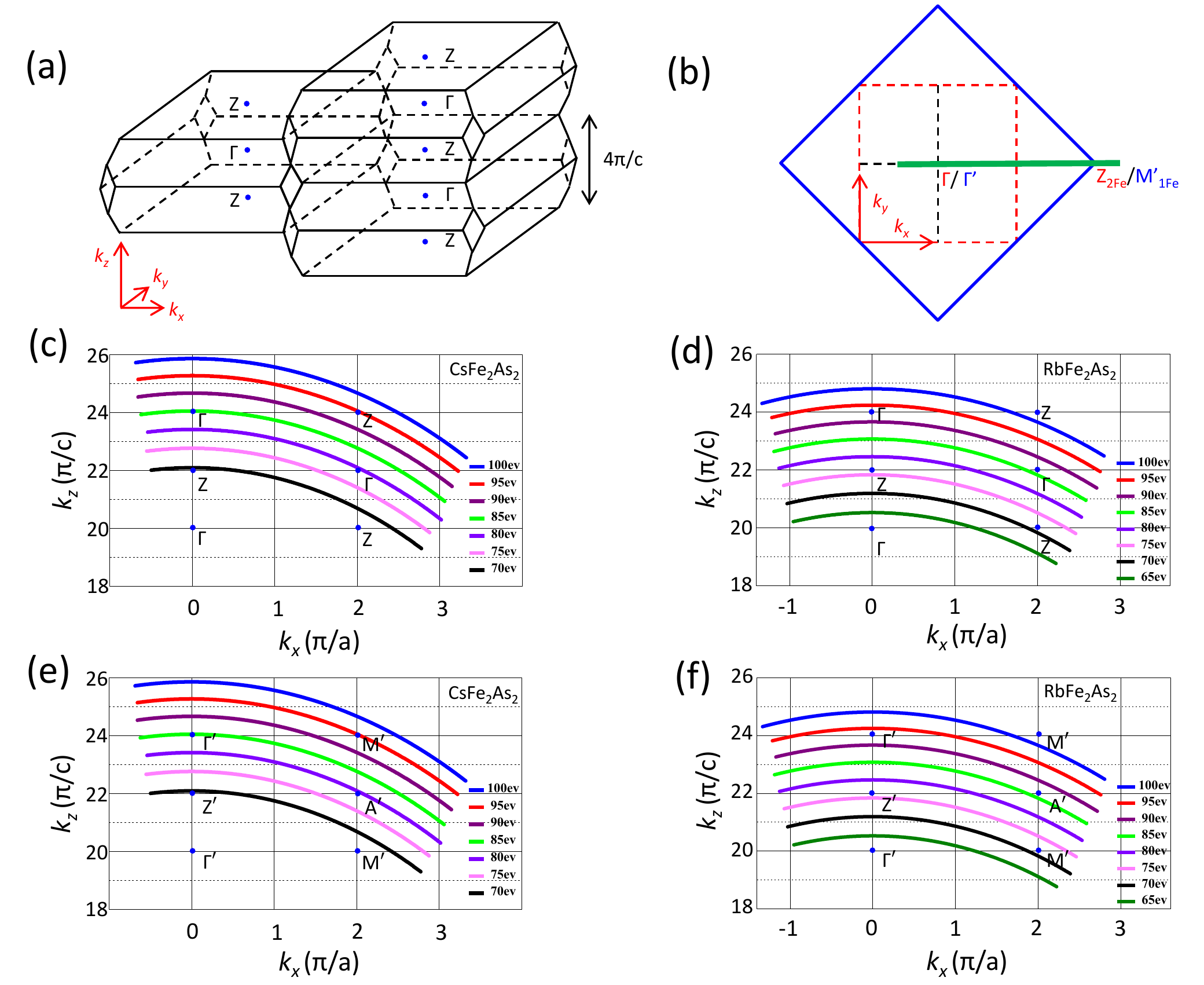}
\caption{(a) The Brillouin zone of 122 iron-based superconductor. (b) A sketch of Brillouin zones of the 2-Fe (red dashed lines) and 1-Fe (blue solid lines) scenarios. (c,d) The calculated momentum cuts for the 2-Fe BZ, along the green line in panel b with various photon energies for CsFe$_{2}$As$_{2}$ and RbFe$_{2}$As$_{2}$, respectively. An inner potential of 15 eV was used for the calculations. (e,f) Same calculations as panels c and d for the 1-Fe BZ.  
} 
\label{fig5}
\end{figure}

For a comparison of $k_z$ dispersion, we have measured the spectral weight distribution around the Fermi levels  in CsFe$_{2}$As$_{2}$ and RbFe$_{2}$As$_{2}$, using various photon energies. The data are shown in Figs. 6 and 7. Here, the blue thick lines show the 1-Fe unit cell of the primary Fe square lattice, and the red dashed lines define the crystallographic two-dimensional unit cell with inequivalent Fe at the corners and the center. One can notice that the Fermi surface topologies do not change significantly, though the spectral weight intensity varies with the change of photon energies. In both samples, the experimental Fermi surface topology consistently shows one hole pocket around the zone center and the flower-like pocket around the corner of the 1-Fe BZ. This behavior suggests a very weak dispersion along the $k_z$ direction, though other ``122" compounds exhibit an alternate arrangement of $\Gamma$ and Z points along the  $k_z$.

\begin{figure}[htbp]
\includegraphics[angle=0, width=0.8 \columnwidth]{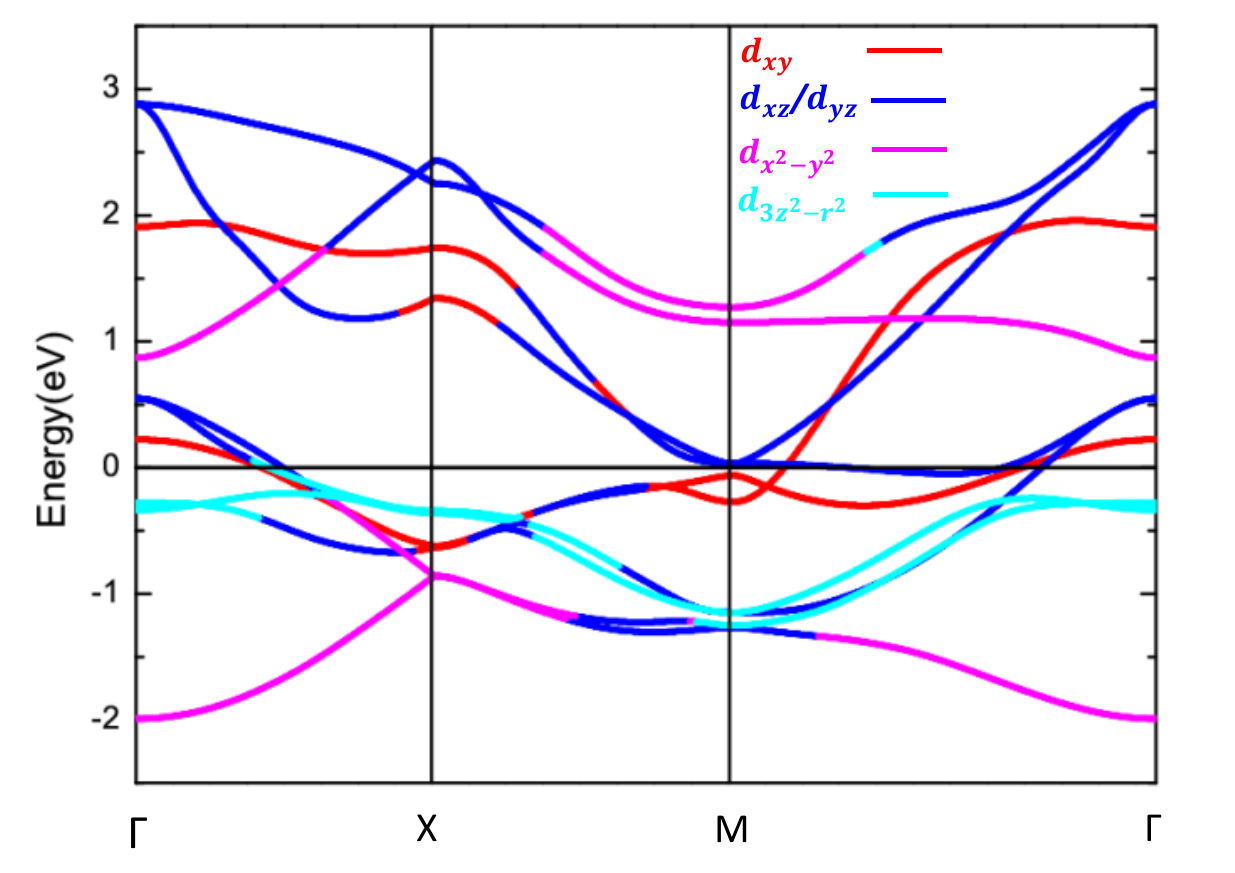}
\caption{Orbital-resolved band structure of CsFe$_{2}$As$_{2}$ plotted along the high symmetry lines for the 2-Fe BZ.}
\end{figure}

\begin{figure}[htbp]\centering
\includegraphics[width=1 \columnwidth]{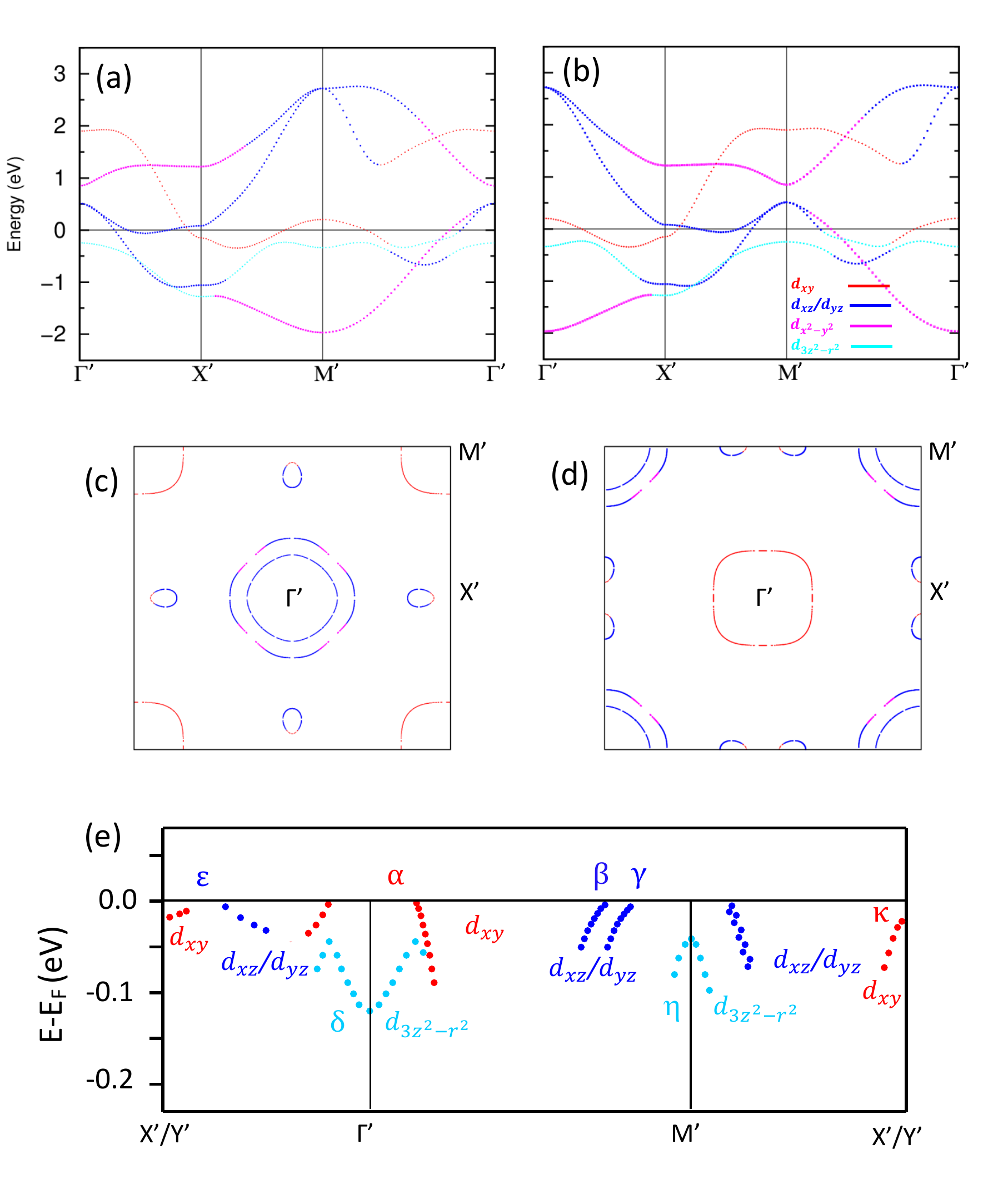}
\caption{(a,b) The unfolded band structures in the 1-Fe BZ and (c,d) the corresponding Fermi surface topologies without and with glide mirror symmetries. (e) Band structure and orbital symmetry resolved from experimental data.
} \label{fig4}
\end{figure}

In Fig. 8, we show how the $\Gamma$-Z cuts and $\Gamma$$'$-M$'$ cuts in Figs. 6 and 7 span in the $k_z$ directions in the 2-Fe and 1-Fe BZs. In the 2-Fe BZ, we expect that the Fermi surface topologies around the $k_x$ = 0 and 2$\pi$/a behave similarly as a function of $k_z$. However, our data in Figs. 6 and 7 show a strong difference between  $k_x$ = 0 and 2$\pi$/a in the 2-Fe BZ. Along the $k_z$ direction in the 1-Fe BZ, Figs. 8(e) and 8(f) show variations of $\Gamma$$'$-Z$'$-$\Gamma$$'$ around the $k_x$ = 0 and M$'$-A$'$-M$'$ around the $k_x$ = 2$\pi$/a, respectively. This arrangement of high-symmetry points in the 1-Fe BZ is in excellent agreement with our data in Figs. 6 and 7.

\begin{figure}[htbp]\centering
\includegraphics[width=1 \columnwidth]{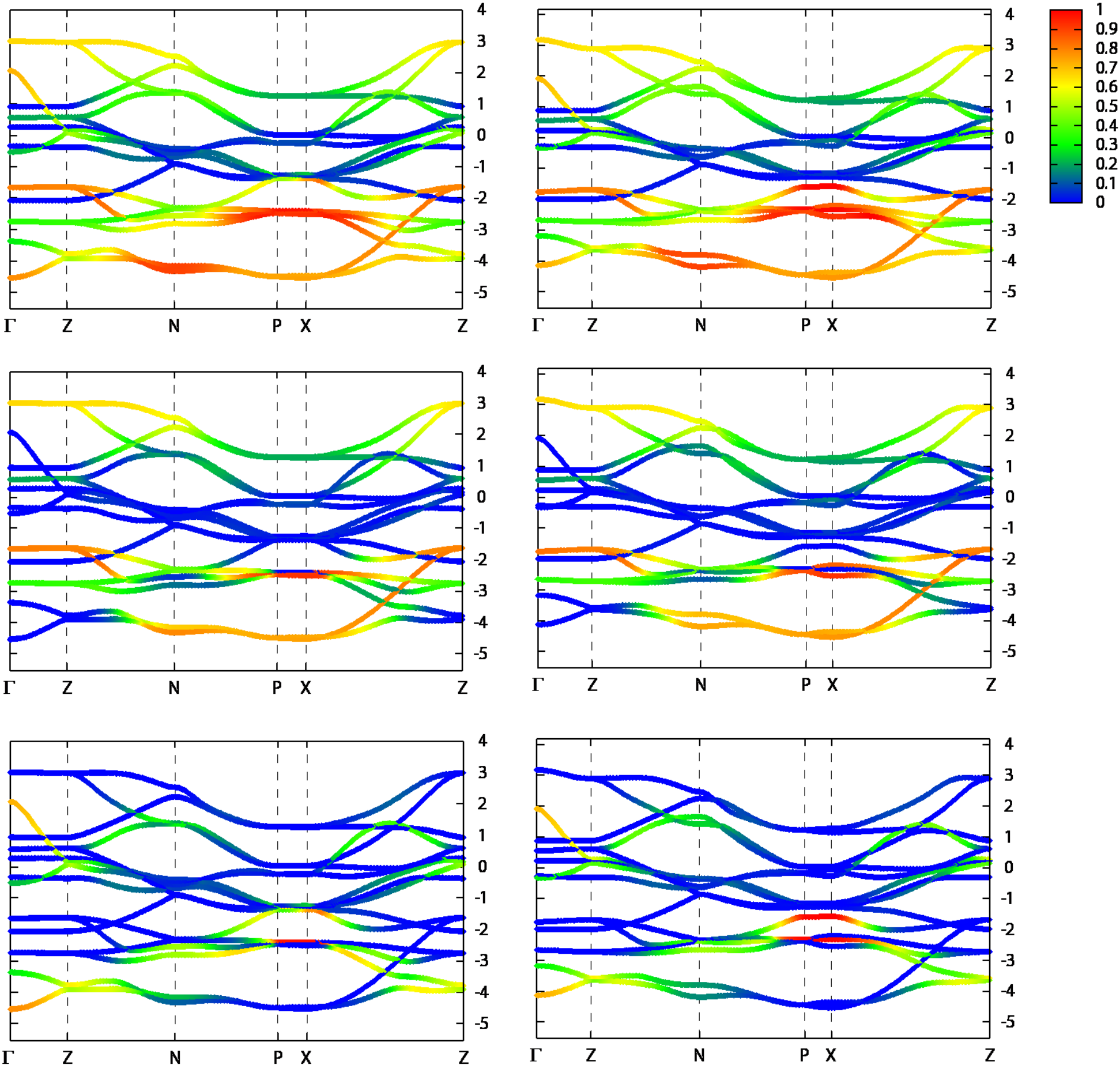}
\caption{As-4$p$ (top panels), 4$p_{x,y}$ (middle panels), 4$p_{z}$ (bottom panels) Wannier orbitals resolved band structure in KFe$_{2}$As$_{2}$ (left panels) and CsFe$_{2}$As$_{2}$ (right panels).}
\end{figure}

\begin{table*}[htbp]
\footnotesize
\caption{Hopping integral parameters $t_{pd}$ between As-4$p$ and Fe-3$d$ Wannier orbitals in CsFe$_{2}$As$_{2}$ in comparison with KFe$_{2}$As$_{2}$. The values of $t_{pd}$ parameters (in units of eV) involved with $z$-component orbitals are bold.}
\label{tab.1}
\begin{center}
\begin{tabular}{lrccccccccccc}
\hline\hline
(K/Cs) &$|t_{pd}|$ &  &$xy$&  & $xz$&  & $yz$&  & $x^{2}-y^{2}$&  &$3z^{2}-r^{2}$\\
\hline & $p_{x}$ &  &0.184/0.22 &  &\bf{0.589/0.589} &  &\bf{0.589/0.589}  &  &0.704/0.711 &  &\bf{0.352/0.307}\\
\hline & $p_{y}$ &  &0.184/0.22 &  &\bf{0.427/0.418} &  &\bf{0.427/0.418} &  &0.704/0.711 &  &\bf{0.352/0.307} \\
\hline & $p_{z}$ &  &\bf{0.604/0.593} &  &\bf{0.086/0.067} &  &\bf{0.086/0.067} &  &  &  &\bf{0.493/0.494}\\
\hline\hline
\end{tabular}
\end{center}
\end{table*}

\textbf{IV. Unfolded Band Structures}

To explore the orbital character of the band structures of CsFe$_{2}$As$_{2}$, we constructed the Wannier functions for the Fe 3$d$ orbitals using the maximally localized Wannier functions (MLWF) scheme, WANNIER90 \cite{CPC178-6851} and WIEN2WANNIER \cite{CPC181-18881}. Fig. 9 shows the Fe 3$d$ orbital-resolved band structures of CsFe$_{2}$As$_{2}$ within the 2-Fe perspective.
In order to make a comparison between the experimental data and band calculations, we attempted to unfold the theoretical band structure into the 1-Fe BZ. Here we have adopted a five-orbital tight-binding model based on the MLWF to approximately unfold the band structures to the 1-Fe BZ. The band dispersions without and with glide mirror symmetries due to the two inequivalent As ions \cite{Anderson1,Brouet1} are shown in Figs. 10(a) and 10(b), respectively. The corresponding Fermi surface topologies are shown in Figs. 10(c) and 10(d).

However, the glide mirror symmetry must be taken into account due to the presence of the As atoms below and above the Fe square lattice in the realistic materials. Thus the unfolded band structure with glide mirror symmetry, {\it i.e.} the second case (Fig. 10(b)), should be adopted to compare with the ARPES data. Indeed, the Fermi surface sheets around the zone center and zone corners are qualitatively consistent with the experimental data, though there are some detailed differences in the Fermi surface topology, which can be attributed to the electronic correlation effects. A major difference between the calculations and ARPES data is the spectral weight distribution around X$'$, which is in agreement with the calculations without the glide mirror symmetry. The reason for this anomaly remains unclear and need further investigations.

\textbf{V. Fe-As coupling in CsFe$_{2}$As$_{2}$}

We argue that the strength of the Fe-As coupling is a crucial ingredient to determine whether the 1-Fe BZ is a realistic choice for the electronic system in CsFe$_{2}$As$_{2}$. If the coupling between Fe and As is weak, in particular along the $z$ direction, the folding potential due to the alternating configuration of As above and below the Fe square lattice will be greatly reduced. In this case, the primary electronic spectral weight of Fe, which dominates the band structure near the Fermi level, will follow the symmetry of the Fe square lattice and reveal the 1-Fe BZ. 

Here we use the hopping integral $t_{pd}$ between Fe-3$d$ and As-4$p$ orbitals as a measure to evaluate the strength of $pd$ hybridization. We constructed the Wannier functions for both Fe-3$d$ and As-4$p$ orbitals (16-band model), and the 4$p$-projected Wannier orbital characters of the As in KFe$_{2}$As$_{2}$ and CsFe$_{2}$As$_{2}$ are plotted in Fig. 11 for a comparison. Along the P-X path, the As-4$p$ spectral weight (mainly $p_{z}$) are entangled with those of Fe-3$d$ in KFe$_{2}$As$_{2}$, while they are disentangled in CsFe$_{2}$As$_{2}$. This difference suggests that the $pd$ hybridization in CsFe$_{2}$As$_{2}$ is considerably weaker than that in KFe$_{2}$As$_{2}$, indicating that the Fe-As coupling is substantially reduced in CsFe$_{2}$As$_{2}$.

In  Table ~\ref{tab.1}, we list the hopping integral parameters $t_{pd}$ between As-4$p$ and Fe-3$d$ Wannier orbitals. It is obvious that the $t_{pd}$ values involved with $z$-component orbitals decrease in CsFe$_{2}$As$_{2}$. The $t_{pd}$ hopping integral between $p_{x}$/$p_{y}$ and $d_{3z^{2}-r^{2}}$ and that between $p_{z}$ and $d_{xz}$/$d_{yz}$ suffer a relatively large decrease. We note here that we neglect the strong electron correlation. If this effect is taken into account, it will further decrease the $pd$ hybridization. Further work is needed to quantitatively describe the effect of the electron correlation. We also notice that the hopping integrals between $p_{x}$/$p_{y}$ and $d_{xy}$/$d_{x^{2}-y^{2}}$ become stronger in CsFe$_{2}$As$_{2}$. 
The weak hopping integrals along the $z$ direction suggests a weak coupling strength between As and Fe in this direction, which can lead to the emergence of the 1-Fe symmetry in the electronic structure.

\end{document}